\documentclass[aps, prl, twocolumn]{revtex4-1}
\usepackage{amsmath}
\usepackage{graphicx}
\usepackage[utf8]{inputenc}
\usepackage{hyperref}

\def\d{\mathrm{d}}
\def\e{\mathrm{e}}
\def\imagi{\mathrm{i}}

\def\kihagy#1{}

\newcommand{\arxiv}[2][]{
  \ifthenelse{\equal{#1}{}}{
    \href{http://arxiv.org/abs/#2}{\texttt{arXiv:#2}}
  }{
    \href{http://arxiv.org/abs/#2}{\texttt{arXiv:#2 [#1]}}
  } 
}

\begin{document}
\title{Weyl discs: theoretical prediction}
\author{Janis Erdmanis}
\author{Árpád Lukács}
\author{Yuli V.~Nazarov}
\affiliation{
Department of Quantum Nanoscience,\\
Kavli Institute of Nanoscience, TU Delft,\\
Lorentzweg 1, 2628CJ Delft, The Netherlands
}

\begin{abstract}
 A  variety of quantum systems exhibit Weyl points in their spectra where two bands cross in a point of three-dimensional parameter space with conical dispersion in the vicinity of the point. 
 We consider theoretically the soft constraint regime where the parameters are dynamical quantum variables. We have shown that in general the soft constraints, in the quasi-classical limit, result in Weyl discs where two states are (almost) degenerate in a finite two-dimensional region of the three-dimensional parameter space.
 We provide concrete calculations for two setups: Weyl point in a four-terminal superconducting structure and a Weyl exciton, i.e., a bound state of Weyl electron and a massive hole. 
\end{abstract}

\maketitle


The Weyl equation is written to describe the propagation of massless fermions \cite{Weyl,Pal}. The $2\times 2$ Weyl Hamiltonian is linear in the particle momenta ${\bf k}$ and has a conical spectrum with degeneracy at ${\bf k} = 0$. The Weyl equation describes neutrini if their masses can be neglected \cite{ChengLi}.

A  variety of quantum systems exhibit similar spectral singularities in the vicinity of crossing of two bands in three-dimensional (3D) parameter space. The degeneracy points are referred to as Weyl points (WP). In solid-state state physics, the parameter space is the Brillouin zone of a crystal lattice and Weyl physics is an active subject in experimental and theoretical research. 
WP are predicted theoretically in Refs.\ \cite{Herring,HasanPred,Murakami}, and have been recently observed experimentally \cite{Soljacicetal,Hasanetal}. For reviews on materials hosting WPs, see Refs.\ \cite{HasanKane, YauFelser}.
In the case of polyatomic molecules, the parameter space for Born-Oppenheimer energy levels is the positions of the nuclei; the existence of points of degeneracy is demonstrated in Refs.\  \cite{herzberg, faure, zhil}. For molecular nanomagnets, the parameter space is the direction and magnitude of the external magnetic field; WPs result  in resonances in tunneling probability \cite{werns1, werns2}.
In the context of quantum transport, a setup with a WP in the space of two gate voltages and a superconducting phases has been proposed to realize a robust quantized current source \cite{leone2008cooper}. WPs have been recently predicted \cite{YNC,Nazarov} in the spectrum of Andreev bound states (ABS) \cite{NBbook} in four terminal superconducting nanostructures where three independent phases form 3D parameter space. Quantized topological transconductance has been predicted. Similarly, WP can be also realized in three terminal nanostructures \cite{MH} and other systems \cite{Zhang1,Zhang2}.

It seems a relevant approximation to treat the parameters forming the space where the WP occurs, as fixed numbers (hard constraint). However, a much more realistic and general situation is where the parameters are dynamical quantum variables, which can be the subject of fluctuations and also back-action from the system hosting the WP. To describe this situation of a \emph{soft constraint}, one would, e.g., promote a parameter $x$ to an operator $\hat{x}$, add an energy term $A(\hat{x}-x_0)^2$ that attempts to constrain $\hat{x}$ to $x_0$ at sufficiently large $A$, and add a Hamiltonian accounting for the dynamics of $\hat{x}$.

\begin{figure}[b!]
  \vspace{-1ex}
 \noindent\hfil\includegraphics[scale=.7]{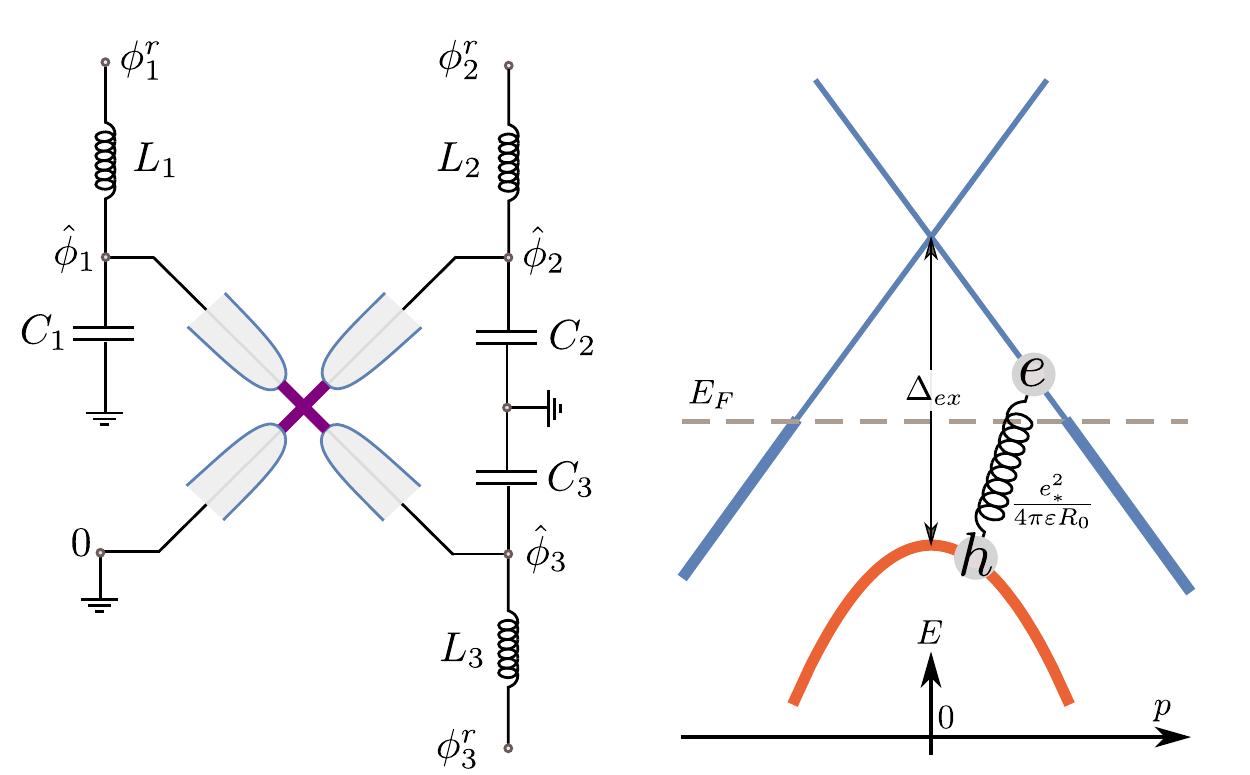}
 \caption{The two setups under consideration. Left: the four-terminal superconducting nanostructure embedded in a linear circuit made of small inductances $L_{1,2,3}$ and capacitances $C_{1,2,3}$. 
 Right: a model band structure that supports Weyl excitons with energy $\approx \Delta_{ex}$ that are bound states of a Weyl electron and a massive hole, with the hole mass providing a soft constraint for the electron momentum.}
 \label{fig:setup}
\end{figure}
In this Rapid Communication, we demonstrate the drastic consequences of a soft constraint in the vicinity of a WP. The degeneracy of two bands that has been restricted to a singular point for a hard constraint, in the quasi-classical limit spreads over to a finite two-dimensional region that we term \emph{Weyl disc}. Quantum effects lift the degeneracy at the disc, resulting in strong anisotropy of the conical spectrum. We assess the situation in detail and provide detailed calculation of the quantum spectrum for two, very different, and physically interesting setups. The first setup is a multi-terminal superconducting nanostructure embedded in a linear circuit. The second setup is an exemplary band structure where a \emph{Weyl exciton} consisting of a Weyl electron and a massive hole can be formed.

Let us shortly stress the relevance of the setup, and the concrete significance of our results; more details are given in Ref.\ \cite{supplmat}. The superconducting nanonstructures (with WPs) can be easily fabricated and implemented as nanodevices, and quantum manipulation in similar devices has been experimentally verified \cite{janvier2015coherent}. The Weyl disc regime described provides extra opportunities for quantum computing owing to the degeneracy of the quantum states, such degeneracies have been the basis of holonomic \cite{ZanardiRasetti, PachosZanardiRasetti} and topological quantum computing \cite{Freedman2008, Sjoqvist, Lahtinen}.
We find practical candidates for Weyl excitons in materials such as graphene, germanene, TaAs, TaP, and NbAs. We predict a unique property of Weyl excitons: In the Weyl disc regime, they can only move in one direction. This can be observed in a simple experiment we describe \cite{supplmat}.

Let us describe the setups in detail. As shown in Ref.\ \cite{YNC}, the ABS spectrum of a four-terminal superconducting nanostructure can have WPs where ABS energy reaches zero (relative to Fermi level). This implies that the ground state of the nanostructure is close to the first excited singlet state. We count the phases from the WP position. The effective Hamiltonian in the vicinity of the WP reads
$\hat{H}_{\rm WP} = (\hbar/2e)I_{na}\hat{\phi}_n \hat{\sigma}_a\,,$
where $\hat{\sigma}_a$ denote the Pauli matrices in the space of ground and excited singlet states \cite{YNC}. The soft constraint situation occurs naturally if one takes into account self-inductances of the superconducting leads and associated capacitances (see Fig.\ \ref{fig:setup}).
This promotes the superconducting phases at the nanostructure to dynamical variables $\hat{\phi}_n$, which are softly constrained to the superconducting phases $\phi^r_n$, fixed by the magnetic fluxes in the corresponding superconducting loops. The full Hamiltonian encompasses inductive and capacitive energy and reads \cite{YNC,AL1,LLO}
\begin{equation}
 \label{eq:H}
 \hat{H} = \hat{H}_{\rm WP} + \sum_n \left[ \frac{(\hbar/2e)^2}{2L_n}(\hat{\phi}_n - \phi^r_n)^2
  + \frac{(2e\hat{N}_n)^2}{2C_n}\right]\,. 
\end{equation}
Here the number operators $\hat{N}_n$ are canonically conjugate variables to the phases $\hat{\phi}_n$: $[\hat{N}_n,\hat{\phi}_m]=-\imagi \delta_{nm}$ \cite{NBbook}. Here the inductive energy provides the soft constraint, and the capacitive energy is responsible for the quantum fluctuations of the phases.

For a complementary example with very different physical content, let us consider a solid exemplary band structure (Fig.\ \ref{fig:setup}b). It comprises an electron band with a WP and a parabolic valence band. To soft-constrain the momentum of the Weyl electron, let us tie it to a massive hole coming from the valence band. The bond is naturally provided by the Coulomb interaction, and the resulting particle is a sort of exciton, described by the Hamiltonian
\begin{equation}
 \label{eq:Hexc}
 \hat{H}_{\rm ex} = \Delta_{\rm ex}+\hat{H}_{\rm WP}+\sum_n  \frac{(\hat{p}_n - p^T_n)^2}{2 m_n^*} - \frac{e_*^2}{4\pi\epsilon_0 r}\,,
\end{equation}
where we count all momenta from the quasi-momentum of the Weyl point, $\hat{H}_{\rm WP}= v_{na}\hat{p}_n\hat{\sigma}_a$,
 $\hat{p}_n$ are the components of the quasi-momentum of the Weyl electron, $p_n^T$ are those of the total exciton quasi-momentum, $m_n^*$ are the (possibly anisotropic) hole masses, and the last term presents Coulomb attraction between electron and hole, with $r = |{\bf r}|$ being the distance between these two particles. 

Let us note the close similarity: $H_{\rm WP}$ and the soft constraint term in Eq.\ (\ref{eq:Hexc}) are brought to the form in Eq.\ (\ref{eq:H}) with the replacements $p_n \to P\phi_n$,   $P v_{na}\to (\hbar/2e) I_{na}$, $(P^2/2 m_n^*) \to (\hbar/2e)^2/2 L_n$, where $P$ is a constant with momentum dimension. Since ${\bf r}$ is canonically conjugate to ${\bf p}$, the Coulomb energy plays a role similar to the capacitive energy in the Hamiltonian (\ref{eq:H}), providing the quantum fluctuations of ${\bf p}$.

For both setups, we evaluate the energies of the discrete quantum states, analyzing their dependence on  the parameters, either $\phi_n^r$ or $p^T_n$.

Systems described by the Hamiltonians (\ref{eq:H}) and (\ref{eq:Hexc}), depending on the parameters, can be in two regimes: the quasi-classical and the opposite, deeply quantum one.

To understand the regimes, let us consider the one-dimensional version of Eq.\ (\ref{eq:H}).
It is exactly solvable, since the quasi-spin part has a single spin component, which can be diagonalized simultaneously with the Hamiltonian. For the spin eigenvalue $\sigma=\pm 1$, the kinetic part of the Hamiltonian is $(\hbar/2e)^2 (1/2L)(\hat{\phi}-\phi^r + \sigma \phi_0)^2 -  LI^2/2$, with $\phi_0 = (2e/\hbar) IL$. At $\phi^r=0$, it gives rise to two degenerate minima separated by $2\phi_0$ with an energy barrier  $E_B=LI^2/2$ between them. The Hamiltonian for both values of $\sigma$ is that of a harmonic oscillator, with frequency $\omega=1/\sqrt{LC}$. The quasi-classical parameter $\mathcal{Q}$ is defined as the ratio of the barrier height and the energy quantization of the oscillators, and reads
\begin{equation}
 \label{eq:qcp}
 \mathcal{Q}
 =\frac{1}{2}\left(\frac{LIe}{\hbar}\right)^2 \frac{\hbar}{e^2Z}\,,
\end{equation}
where $Z=\sqrt{L/C}$ is the characteristic impedance of the oscillator. In Eq.\ (\ref{eq:qcp}), an estimation for the first term is ratio of the inductance of the circuit to the typical inductance of the nanostructure, which has to be small to provide good confinement. However, the second term is large, estimated as the ratio of vacuum impedance to resistance quantum $\sim 10^2$. This is why the quasi-classical limit $\mathcal{Q}\gg 1$ is well achievable (see detailed estimations in \cite{supplmat}).
In a 3D case, we define $\mathcal{Q}$ with respect to the maximal $L_n I_n^2$ (\emph{easy direction}).

Similar analysis for the Hamiltonian (\ref{eq:Hexc}) yields in one dimension (1D) a barrier height of $E_B=m^* v^2/2$. The parameter $\mathcal{Q}$ is defined as the ratio of the barrier height to the ground state Coulomb binding energy $E_b \propto (e_*^2/4\pi\epsilon_0)^2 m^*/2\hbar^2$,
yielding
\begin{equation}
 \label{eq:qcpE}
 \mathcal{Q} = \left( \frac{\hbar v 4\pi\epsilon_0}{e_*^2} \right)^2\,.
\end{equation}
If one estimates the Weyl velocity $v$ with the typical Fermi velocity for metals $v_{F}\sim 10^6 {\rm m} {\rm s}^{-1}$, and the dielectric constant as $\epsilon_r\approx 10$, $\mathcal{Q}\sim 25$, the quasi-classical limit is well achievable in solids.
In a 3D case, we define $\mathcal{Q}$ with the parameters in the easy direction (maximal $m v^2$).

The deeply quantum limit $\mathcal{Q}\ll 1$ is in fact not interesting, since there the Weyl energy is not modified by the soft constraint, except for trivial  perturbative corrections.

In this study, we concentrate on the quasi-classical limit. We give analytical results valid at $\mathcal{Q}\gg 1$ and numerical results for $\mathcal{Q}\sim 5$.

\begin{figure}
\noindent\hfil\includegraphics{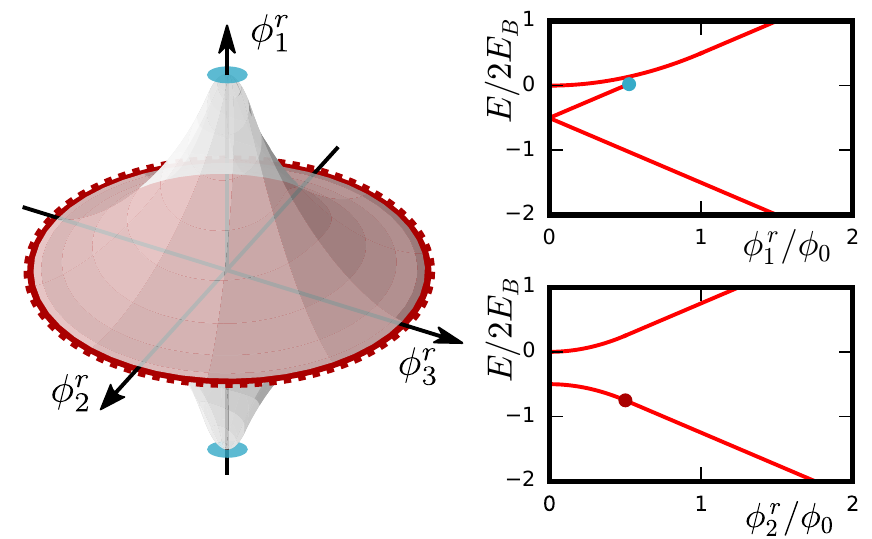}
\caption{The Weyl disc. \emph{Left}: The region in parameter space $\phi^r_n$ (or $p^T_n$) where three quasi-classical energy minima exists. The two minima are degenerate at the disc in the plane $(\phi^r_2,\phi^r_3)$. \emph{Right}: The quasi-classical energy spectrum in the easy direction (\emph{top}) and in a direction within the disc (\emph{bottom}). The dots mark the region edges. (The parameter choice is $L_n=L/n$ and $I_n=I$. )
 }
 \label{fig:qcl}
\end{figure}

In the quasi-classical regime, we neglect the fluctuations of the phases $\phi_n$ and replace the quasi-spin term $H_{\rm WP}$ with one of its eigenvalues.
The matrix $I_{na}$ can be diagonalized by a coordinate transformation $I_{na}\to I_n \delta_{na}$ \cite{supplmat}. Then we need to minimize
\begin{equation}
 \label{eq:EclSdiag}
 E_{{\rm cl},\sigma} = \frac{\sigma\hbar}{2 e} \sqrt{\sum_n I_n^2 \phi_n^2} + \left(\frac{\hbar}{2e}\right)^2 \sum_n\frac{(\phi_n-\phi_n^r)^2}{2L_n}\,.
\end{equation}
If $|\phi_n^r|\gg \phi_0$, the minimization reproduces the two cones of the Weyl spectrum, $(\sigma\hbar/2e)\sqrt{\sum_n I_n^2 (\phi_n^r)^2}$.
In the vicinity of the Weyl point $|\phi_n| \sim \phi_0$, the Weyl spectrum is drastically modified (see Fig.\ \ref{fig:qcl}). Most importantly, the minimization gives two minima for $\sigma=-1$ in the 3D region shown in the figure. These two minima are precisely degenerate at a 2D \emph{Weyl disc},
which is perpendicular to the \emph{easy direction}, where $L_nI^2_n$ is maximal ($n=1$ for the easy direction). The disc is an ellipse with dimensions $(4e/\hbar)(L_1 I_1^2 - L_m I_m^2)/I_m$, $m=2,3$.

In Fig.\ \ref{fig:qcl}, we plot the energies along the easy direction and in the plane of the disc. There is a linear dependence of the energies in the easy direction. The second minimum for $\sigma=-1$ disappears at a critical value of $\phi^r_1$. For even larger $\phi_1^r$, the Weyl spectrum
$E\approx(\hbar/2e) I_1 \sigma \phi_1^r$ is seen again. If we move along the disc, two minima remain degenerate until they merge at the disc edge. 

The same minimization applies to the Weyl exciton setup. In this case, the lowest curves in Fig.\ \ref{fig:qcl} define the lower boundary of the continuous spectrum. The bound exciton states follow the edge at slightly lower energy, with binding energy $E_b \ll E_B$. If we move along the disc, all bound states remain doubly degenerate, until the edge of the disc. They split linearly if we move in the easy direction.

This brings us to the main conclusion of the paper:  In the quasi-classical limit $\mathcal{Q}\gg 1$, soft constraints extend the isolated degeneracy in the WP into a finite 2D region. This property of WP can be used for the purposes of quantum manipulation and computation.

\begin{figure}
  \noindent\hfil\includegraphics{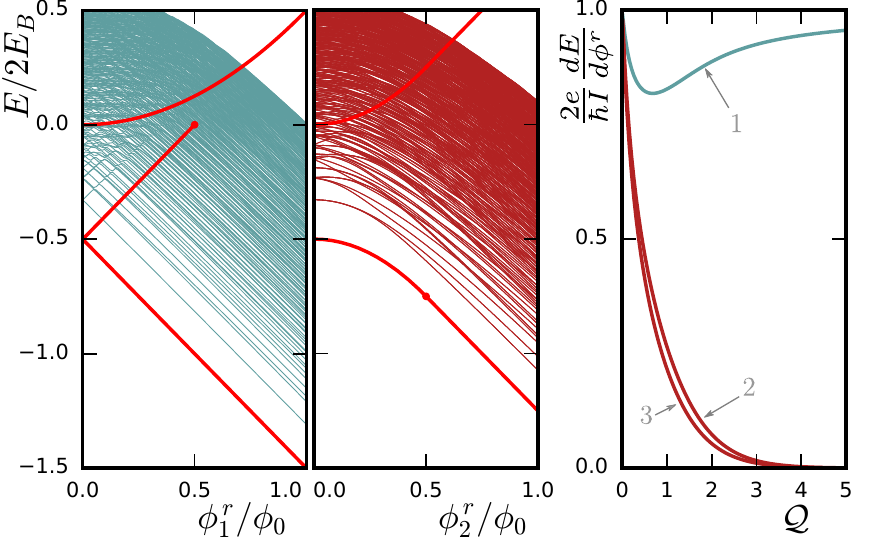}
\caption{
  The energy spectrum of the circuit shown in Fig.\ \ref{fig:setup} for $\mathcal{Q}=5$ in the easy direction (\emph{left panel}) and in the plane of the disc (\emph{center panel}). The parameters are
  $L_n=L/n$, $C_n=C=\hbar^2 \mathcal{Q}^2/L E_B^2$, $I_n=I$, and $\mathcal{Q}=5$ ($L$ and $I$ arbitrary).
  We also show the ``velocities'' $\partial E/\partial\phi_n^r$ versus the quasi-classical parameter $\mathcal{Q}$ 
  in the ground state
  (\emph{right panel}).
  }
 \label{fig:junction}
\end{figure}

At large but finite values of $\mathcal{Q}$, the degeneracy at the disc is lifted, albeit the corresponding energy splittings remain relatively small at moderate values of $\mathcal{Q}$. We illustrate this with numerical results for both setups. In Fig.\ \ref{fig:junction}, we plot the full energy spectrum of the superconducting nanostructure for $\mathcal{Q}=5$. Besides the ground state, the spectrum includes the corresponding excitations in three oscillators.
For comparison, in Fig.\ \ref{fig:junction}, we plot in red the quasi-classical results from Fig.\ \ref{fig:qcl}. Upon a small shift, the lowest curves give good approximations for the numerical energies of the lowest states.
At $\phi^r=0$, all levels are doubly degenerate. If we move in the easy direction, the levels are split with $\Delta E \propto (\hbar/e)I_1 \phi^r_1$. The levels become increasingly dense at higher energies. Since the levels begin to cross, this behavior is restricted to increasingly small values of $\phi^r_1$. At $\phi_1^r < 0.5 \phi_0$,  the crossings are avoided at an exponentially small energy scale corresponding to the tunneling amplitude between the minima.
The amplitude increases with energy owing to a bigger overlap of the oscillator excited states in two minima.

If we move in a perpendicular direction, we observe an exponentially small energy splitting at $\phi_{2,3} \approx 0.4 \phi_0$. At small $\phi^r_{2,3}$, the splitting is $\Delta E \approx (\hbar/e) I_{2,3}\phi^r_{2,3}\e^{-2\mathcal{Q}}$ in the ground state \cite{supplmat}. We see this suppression in the plot of the normalized ``velocities'' of the lowest state, $(2e/\hbar I_1)\partial E/\partial \phi^r_n$ at $\phi^r\to 0$ (Fig.\ \ref{fig:junction}, right panel). In the deep quantum limit, $\mathcal{Q}\lesssim 1$, all velocities remain the same as for the original Weyl spectrum. The velocity in the easy direction stays closer to this value at any $\mathcal{Q}$.

\begin{figure}
 \noindent\hfil\includegraphics[scale=1]{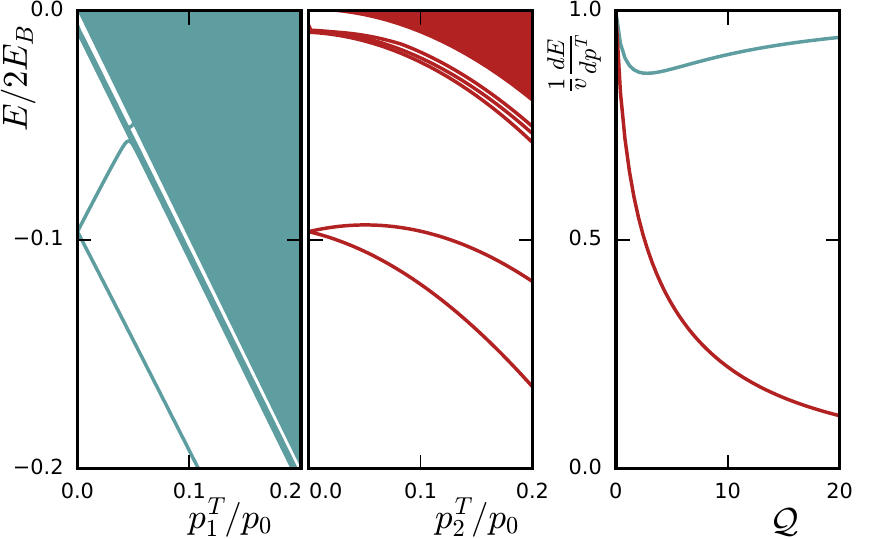}
 \caption{
 The energy spectrum of a two-dimensional anisotropic Weyl exciton for $\mathcal{Q}=20$ in the easy direction (\emph{left panel}) and in the plane of the disc (\emph{center panel}). The parameters are $m_n=m/n$, $v_n=v$, $e_*^2=\hbar v 4\pi\epsilon_0/\sqrt{\mathcal{Q}}$ and $\mathcal{Q}=20$ ($v$ and $m$ arbitrary). We also show the velocities $\partial E/\partial p^T_n$ versus the quasi-classical parameter $\mathcal{Q}$ 
 in the lowest state of the exciton
 (\emph{right panel}).
 }
 \label{fig:exciton}
\end{figure}

In Fig.\ \ref{fig:exciton}, we show the spectrum of the exciton Hamiltonian (\ref{eq:Hexc}) for $\mathcal{Q}=20$. For the sake of numerical efficiency, we have computed the spectrum in 2D limit. This is valid in the highly anisotropic limit $m^*_3 \ll m_{1,2}^*$. Also, graphene provides a practical example of a stable conical spectrum in 2D. 
With graphene data, $v\approx v_F$ and a substrate with a relative permittivity $\sim 10$, $\mathcal{Q}\sim 20$ \cite{ElPGrap}.
The continuous spectrum is shown by the shaded region. Its lower edge is given by the quasi-classical result (Fig.\ \ref{fig:qcl}). Below the edge, we plot the energies of the five lowest bound states. 
If we go in the easy direction, we observe an almost unmodified Weyl spectrum for the lowest and the first excited states. In contrast to this, the splitting between these states remains small in the plane of the disc. This is seen for the lowest and the first excited states as for third and fourth excited states that are close to the edge. In the right panel of Fig.\ \ref{fig:exciton}, we plot the normalized velocities of the lowest state versus $\mathcal{Q}$. Similarly to the case of the superconducting nanostructure, the Weyl velocity in the easy direction is hardly modified, while that in perpendicular direction is strongly suppressed with increasing $\mathcal{Q}$.
In fact, the wave function of the bound state near one of the minima is singular in coordinate space owing to the the singularity of the Coulomb potential at $r\to 0$. The calculation of the amplitude of tunneling  between the minima demonstrates that the value of the amplitude is determined by this singularity. This results in power-law suppression $\partial E/\partial p_n^T = \pm v_n /2\mathcal{Q}^4$ in the ground state \cite{supplmat} in 3D. In 2D, $\partial E/\partial p_n^T = \pm v_n /2(\mathcal{Q}/4)^3$. [In 2D, we use the definition $\mathcal{Q}=4 E_B/E_b$ to obtain Eq.\ (\ref{eq:qcpE}).]

In conclusion, we have shown that a Weyl spectrum is essentially modified by soft constraints of the spectral parameters in the quasi-classical limit. A Weyl disc emerges in the vicinity of the WP. There are two degenerate states at the disc, that are slightly split at moderate values of the quasi-classical parameter $\mathcal{Q}$. 

We illustrate this general statement with two examples of very different physical systems. The first system is a multi-terminal superconducting nanostructure where the spectral parameters are the superconducting phases and the soft constraint is realized by an external circuit. The second example concerns a Weyl exciton that is the bound state of a Weyl electron and a massive hole. The mass provides a soft constraint of the total exciton quasi-momentum to the momentum of the Weyl electron. We show that in both examples, the quasi-classical regime can be achieved with a reasonable parameter choice.

This project has received funding from the European Research Council (ERC) under the European Union's Horizon 2020 research and innovation programme (Grant Agreement No.\ 694272).

\bibliography{wdp}

\newpage
\appendix
\section{Supplementary Material}\label{app:supplmat}
In this supplementary material, we present the details of the discussion and the detailed derivations of the results in the main text.
We explain the possible use of Weyl disks in the context of quantum manipulation. We provide concrete examples of realizations of Weyl excitons in various materials with detailed estimation of parameters and propose a decisive experiment to reveal their unusual properties.
We present the detailed minimization of the energy in the quasi-classical limit for both setups considered in the main text:  the superconducting nanostructure and the Weyl exciton.
We present the perturbation theory for the splitting at the Weyl disc, assuming a large but finite quasi-classical parameter $\mathcal{Q}$. 
We shortly summarize the numerical approach in use.

\subsection{Weyl disks for quantum computing applications}

The coherent manipulation of the Andreev states in superconducting nanostructures, and their readout, has been reported in  Ref.\ \cite{janvier2015coherent} and references therein. It was a resonance manipulation whereby a high-frequency modulation of the superconducting phase difference with the frequency matching the splitting between the singlet states was used to create the superposition of these two states. Rather long coherence times have been measured. The advantage of superconducting structure is that it is an integral part of an electric circuit and the manipulation
can be achieved by an electric signal.

A four-terminal superconducting nanostructure can be brought by tuning the three superconducting phases to the Weyl disk parameter range. In this range, the singlet states are degenerate in a 2D submanifold of 3D parameter space. Such degeneracies open up unique  possibilities in quantum computing being the physical basis of holonomic \cite{ZanardiRasetti, PachosZanardiRasetti} and \cite{Freedman2008} topological quantum computing (see Refs.\ \cite{Sjoqvist, Lahtinen} for reviews.) A close physical realization for superconducting nanostructures are Majorana states in semiconducting nanowires \cite{Majorana} The coherent manipulation in degenerate subspaces can be achieved by adiabatic changes of the parameters(see Ref.\ \cite{NBbook}): in our case, the superconducting phases. An advantage of such schemes is the protection  against parametric noise \cite{SolinasZanardiZanghi} and thus enhanced coherence times. 

For Weyl disks, the protection is not absolute since the states can be split in energy by a change of one of the three phases. However, this enables the schemes that combine adiabatic and resonant manipulation and cannot be realized in more protective setups.

\subsection{Candidate realizations of Weyl excitons}
In the main text, we have considered a model of an exciton composed of a quasiparticle with a conical spectrum and a quasiparticle of opposite charge with a parabolic dispersion, that is, a massive one. To our knowledge, such excitons have not yet been observed. Here we present several candidate realizations that we have found in various materials. The list of the candidates is not exhaustive.

In Table \ref{tab:mat} we summarize the data of these materials, as well as the references to the sources of the data. In Table \ref{tab:exdat}, we give the computed exciton parameters as it follows from the quasi-classical approximation. The binding energy was evaluated as $E_b =(e^2/4 \pi \epsilon_0 \epsilon)^2 m^*/2 \hbar^2$ and quasi-classical parameter as $\mathcal{Q} = ( \hbar v  4 \pi \epsilon_0 \epsilon/e^2)^2$ [see Eq.\ (\ref{eq:qcpE})]. For 2D materials, $\epsilon_r$ is the relative effective dielectric constant of the substrate: mounting the material on different substrates permits control over the electron-hole interaction and thus for tuning the binding energy. All candidate realizations considered show a big quasi-classical parameter $\mathcal{Q}$.

The most famous realization of conical spectrum is found in 2D materials of graphene class. Strictly speaking, the conical point is no Weyl point since the Weyl point can be realized in three dimensions only. However, the structure of the Hamiltonian is the same and all the results obtained in the main text are valid. The conical point is fixed to K point of the Brillouin zone. The closest band with parabolic maximum in K point is at energy distance $10\, {\rm eV}$ for graphene \cite{MAH, KoNa, w} and $4\, {\rm eV}$ for much more exotic germanene \cite{MAH, germ} that is a 2D layer of germanium atoms. This band is rather flat so the corresponding particles are massive, $m^* \simeq 20 m_e$.

 The conical point in the exciton spectrum is at zero total momentum. Thus the excitons can be produced by (far) ultraviolet radiation of proper frequency. Changing the incident angle and frequency of the radiation, one can selectively excite the particles with fixed wave vector and thus verify the dispersion relation. The symmetry of K point implies same masses in both crystal directions, so in default these excitons do not show Weyl disk behavior. However, the asymmetry of the mass tensor can be achieved by uniaxial deformation of the material or mounting it on an anisotropic substrate.  

True Weyl points are found in 3D crystals TaAs, TaP and NbAs \cite{GPMC, BJCS, LXHS}. Their locations in the Brillouin zone are not pinned to any symmetry points.
The closest parabolic maximum is found at Z symmetry point. The corresponding gaps are small so excitons are excited by infrared radiation. The exciton energy minimum is not at zero wave vector but rather at $k_0 = 0.2\, {\rm nm}^{-1}$. Although this wave vector is short in comparison with the inverse lattice constants, the relatively long wavelength of the radiation makes it difficult to directly produce these excitons. However, they, as any other excitons with non-zero wave vector, can originate from the processes that involve an emission of a phonon \cite{K2,loudon1962theory}, in this case, the excess $k_0$ is given to a phonon. An alternative method would involve the periodic spatial modulation of the dielectric constant at the surface with the corresponding period of $35$ nm. 

High dielectric constants of the materials result in low binding energies and large values of the quasi-classical parameter. Due to this, and the strong anisotropy of the mass tensor at Z point, we expect almost perfectly degenerate Weyl disks.

\begin{table}
\begin{center}
\begin{tabular}{c||c|c|c|c|c|c|c|c}
  Material & Ref.                    & $a,c$     & PPB      & $\Delta p$    & $\Delta_{\rm ex}$ & $m_*$ & $v$                        & $\epsilon_r$\\
           &                         & \AA       &          & $\hbar \pi/a$ & eV                & $m_e$ & $10^5 {\rm m}{\rm s}^{-1}$ & \\
 \hline\hline
 graphene  & \cite{MAH, KoNa, w}     & 2.5       & K        & 0             & 10                & 20    & 6                          & $17\dots 37$\\
  germanene & \cite{MAH, germ}        & 3.8       & K        & 0             & 4                 & 20    & 3                          & $12\dots 37$\\
  TaP      & \cite{GPMC, LXHS}       & 3.4,11.5  & $\Sigma$ & 0.02          & 0.3               & 0.2   & 4                          & 190\\
 TaAs      & \cite{GPMC, BJCS, LXHS} & 3.5,11.8  & $\Sigma$ & 0.02          & 0.4               & 0.2   & 4                          & 100\\
  NbAs      & \cite{GPMC, LXHS}       & 3.5,11.8  & $\Sigma$ & 0.02          & 0.1               & 0.2   & 4                          & 250\\
\end{tabular}
\end{center}
\caption{Data of some materials with Weyl points in their band structure. Notation (other than those used in the main text): $a$ and $c$ are the lattice constants, $\Delta p$ the distance between the Weyl point and the parabolic minimum in the Brillouin zone and PPB the position of the parabolic minimum in the Brillouin zone.}
 \label{tab:mat}
\end{table}

\begin{table}
\begin{center}
\begin{tabular}{c||c|c}
 Material  & $E_b$            & $\mathcal{Q}$\\
           & eV               &              \\
 \hline\hline
 graphene  & $1.0\dots 0.2$   & $21\dots 103$\\
  germanene & $2.0\dots 0.2$   & $3\dots 26$\\
  TaP      & $1\cdot 10^{-5}$ & 1000 \\
 TaAs      & $3\cdot 10^{-4}$ & 300\\
 NbAs      & $5\cdot 10^{-5}$ & 1800\\
\end{tabular}
\end{center}
\caption{Data of Weyl excitons calculated from material data in Table \ref{tab:mat}.}
 \label{tab:exdat}
\end{table}

\subsection{Possible experiments with Weyl excitons}
The most surprising and remarkable property 
of Weyl excitons with wave vectors near the Weyl disk is that they can propagate only in two opposite directions parallel to the easy axis,
since their energy does not depend on the wave vector components perpendicular to the easy axis. 

\begin{figure}
 \noindent\hfil\includegraphics{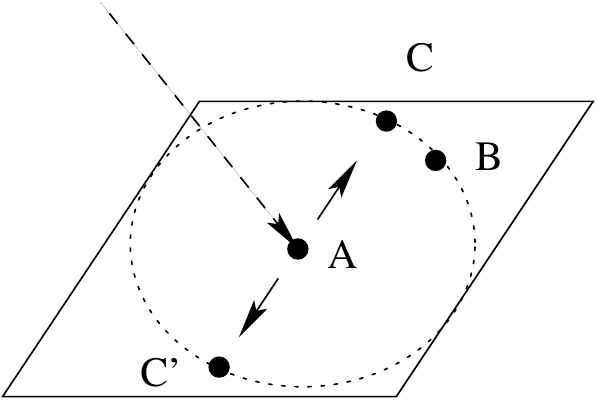}
 \caption{A proposed experiment for the detection of Weyl excitons. The excitons are created at the point A. The luminescent radiation comes from the line $C'-A-C$}
 \label{fig:exp}
\end{figure}

This can be confirmed in the following simple experiment (Fig.\ \ref{fig:exp}). We consider a crystal sample with a surface normal vector perpendicular to the main axis (shown by double arrow).  The Weyl excitons are created near the point A in the sample, in a spot of the incident radiation. If the excitons were usual, we expect them to fly away from the point A, scatter in all directions and finally decay producing the weak luminescent radiation in the whole sample volume. In  contrast to this, the Weyl excitons propagate along a line parallel to the easy axis and remain near the sample surface even if scattered. Therefore we expect intense luminescence, for instance, from points C and C', but not from the point B.

\subsection{Quasi-classical approximation: nanostructure}\label{sec:Sqcl}
The quasi-classical energy expression is obtained from the Hamiltonian (\ref{eq:H}) by replacing the operator $\hat{\phi}_n$ with its mean value $\phi_n$, and at the same time, the quasi-spin part of the Hamiltonian $H_{\rm WP}$ with one of its eigenvalues $\sigma \sqrt{I_{na}I_{ma}\phi_n \phi_m}$, with $\sigma=\pm 1$. The resulting expression is
\begin{equation}\label{eq:SEcl}
 E_{{\rm cl},\sigma} = \left(\frac{\hbar}{2e}\right)^2 \sum_n \frac{(\phi_n-\phi_n^r)^2}{2L_n} + \frac{\sigma\hbar}{2e}\sqrt{I_{na}I_{ma}\phi_n\phi_m}\,.
\end{equation}
Note, that it $\phi = I^{-1}O I_0 \phi'$, where $I_0={\rm diag}(I_1,I_2,I_3)$, and $O=(O^T)^{-1}$ is such that $G^{-1} = I_0^{-2}O I^T L I O^T$ is diagonal, then the form (\ref{eq:EclSdiag}) is achieved.

To obtain the quasi-classical approximation of the energy levels, one shall minimize (\ref{eq:SEcl}) w.r.t.\ $\phi_n$. This amounts to either setting $\phi_n=0$ (minimum at the boundary), or solving the equation
\begin{equation}
 \label{eq:SEclD}
 \frac{\partial E_{{\rm cl},\sigma}}{\partial \phi_n} = \left(\frac{\hbar}{2e}\right)^2 \frac{\phi_n-\phi_n^r}{L_n} +
  \frac{\sigma\hbar}{2e}\frac{I_n^2\phi_n}{\sqrt{\sum_k I_k^2\phi_k^2}}=0
\end{equation}
and verifying that the second derivative matrix,
\begin{equation}
 \label{eq:SEclDD}\begin{aligned}
 \frac{\partial^2 E_{{\rm cl},\sigma}}{\partial\phi_n\partial\phi_m} =
  &\left(\frac{\hbar}{2e}\right)^2 \frac{\delta_{nm}}{L_n}\\
  &+ \frac{\sigma\hbar}{2e}\left[\frac{I_n^2\delta_{nm}}{\sqrt{\sum_k I_k^2\phi_k^2}}
  + \frac{I_n^2\phi_n I_m^2\phi_m}{\left( \sum_k I_k^2\phi_k^2\right)^{3/2}}
  \right]
  \end{aligned}
\end{equation}
is positive definite. The boundary of the solid body, where three solutions (one for $\sigma=1$ and two for $\sigma=-1$) exist in the left hand side of Fig.\ \ref{fig:qcl} is given by the vanishing of one eigenvalue (and hence the determinant) of the matrix (\ref{eq:SEclDD}). There, one of the $\sigma=-1$ solutions ceases to be a true minimum.

For simplicity sake, let us assume in what follows, that $0 < L_3 I_3^2 < L_2 I_2^2 < L_1 I_1^2$. The direction corresponding to the latter one, $\phi_1$ is thus the easy direction.

The minima of $E_{{\rm cl},\sigma}$ are obtained as follows:
\begin{enumerate}
 \item For $\sigma =1$ and $\sum_n(\hbar/2e)^2 (\phi_n^r/L_n I_n)^2  < 1$, the minimum is at $\phi_n=0$, as there is no solution to Eq.\ (\ref{eq:SEclD}).
 Here, the obtained minimum is $E_{{\rm cl},+} = \sum_n(\hbar/2e)^2 (\phi_n^r)^2/2L_n$.
 
 \item For $\sigma =1$ and $\sum_n(\hbar/2e)^2 (\phi_n^r/L_n I_n)^2  \ge 1$, the solutions are obtained in a parametric form as
 \begin{equation}
  \label{eq:Sparam}
  I_n \phi_n = r u_n\,,\quad I_n \phi_n^r = \left( r + \frac{\sigma 2e}{\hbar}L_n I_n^2\right) u_n\,,
 \end{equation}
 where $u_n u_n=1$ (3d unit vector).
 Here, the quasi-classical energy is
 \begin{equation}
  \label{eq:SEclM}
  E_{{\rm cl},\sigma} = \sum_n \frac{L_n I_n^2 u_n^2}{2} + \frac{\sigma\hbar}{2e}r\,.
 \end{equation}
 Items 1.\ and 2.\ shall be referred to as the \emph{upper energy surface}.

 \item For $\sigma=-1$ and $(2e/\hbar )L_2 I_2^2 < r < (2e/\hbar ) L_1 I^2_1$, for a domain of parameters $\phi_n^r$, Eq.\ (\ref{eq:Sparam}) gives minima (\ref{eq:SEclM}). One boundary of this domain is at $r=(2e/\hbar)L_1 I_1^2$, and the other one is determined by the radius $r_c(u)$, where the second derivative matrix (\ref{eq:SEclDD}) ceases to be positive definite (its determinant crosses zero). This surface is shown in Fig.\ \ref{fig:qcl}.
 
 \item For $\sigma=-1$ and $r=(2e/\hbar)L_1 I_1^2$, the \emph{Weyl disc} is obtained. Here $\phi_1^r=0$, and
 \begin{equation}
  \label{eq:SeclWD}\begin{aligned}
  E_{{\rm cl},-} = &-\frac{L_1 I_1^2}{2}\\ &- \left(\frac{\hbar}{2e}\right)^2 \sum_{n\ne 1}\frac{(\phi_n^r)^2/(2L_nL_1)}{(I_1/I_n)^2/L_n - 1/L_1}\,.
  \end{aligned}
 \end{equation}

 \item For $\sigma=-1$ and $r>(2e/\hbar)L_1 I_1^2$, Eq.\ (\ref{eq:Sparam}) yields the minima (\ref{eq:SEclM}).
 
 Items 4.\ and 5.\ shall be referred to as the \emph{intermediate} and the \emph{lower energy surfaces}, respectively.
\end{enumerate}

\subsection{Quasi-classical approximation for the exciton}\label{sec:SqclE}
Let us perform a similar analysis for the Hamiltonian (\ref{eq:Hexc}). The quasi-classical energy is in this case
\begin{equation}
 \label{eq:SEclE}
 E_{{\rm cl},\sigma} = \sum_n \frac{(p_n-p^T_n)^2}{2m_n^*} + \sigma \sqrt{v_{na}v_{ma}p_np_m}\,,
\end{equation}
where, similarly to the case of the nanostructure, with the transformation ${\bf p} \to v^{-1}Ov_0{{\bf p}}'$, where $v_0$ and $m^{-1}=v_0^{-2}Ov^TmvO^T$ are diagonal,
$O^TO=1$, can be transformed to the form
\begin{equation}
 \label{eq:SEclEd}
 E_{{\rm cl},\sigma} = \sum_n \frac{(p_n-p^T_n)^2}{2m_n^*} + \sigma \sqrt{\sum_nv_{n}^2p_n^2}\,.
\end{equation}
The minimization is done in a similar fashion. The first and second derivatives are
\begin{equation}
 \label{eq:SEclED}
 \frac{\partial E_{{\rm cl},\sigma}}{\partial p_n} = \frac{p_n-p^T_n}{m_n^*} + \sigma\frac{v_n^2 p_n}{\sqrt{\sum_kv_{k}^2p_k^2}}\,,
\end{equation}
and
\begin{equation}
 \label{eq:SEclEDD}
 \frac{\partial^2 E_{{\rm cl},\sigma}}{\partial p_n\partial p_m} =
  \frac{\delta_{nm}}{m_n^*}
  + \sigma\left[\frac{v_n^2\delta_{nm}}{\sqrt{\sum_k v_k^2 p_k^2}}
  + \frac{v_n^2 p_n v_m^2 p_m}{\left( \sum_k v_k^2 p_k^2\right)^{3/2}}\,.
  \right]
\end{equation}
The minima are as follows:

\begin{enumerate}
 \item For $\sigma =1$ and $\sum_n(p_n^T/m_n^* v_n)^2  < 1$, the minimum is at $p_n=0$, as there is no solution to Eq.\ (\ref{eq:SEclED}).
 Here, the obtained minimum is $E_{{\rm cl},+} = \sum_n(p_n^T)^2/2m_n^*$.
 
 \item For $\sigma =1$ and $\sum_n(p_n^r/m_n^* v_n)^2  \ge 1$, the solutions are obtained in a parametric form as
 \begin{equation}
  \label{eq:SparamE}
  v_n p_n = s u_n\,,\quad v_n p_n^T = \left( s + m_n^* v_n^2\right) u_n\,,
 \end{equation}
 where $u_n u_n=1$ (3d unit vector).
 Here, the quasi-classical energy is
 \begin{equation}
  \label{eq:SEclEM}
  E_{{\rm cl},\sigma} = \sum_n \frac{m_n v_n^2 u_n^2}{2} + \sigma s\,.
 \end{equation}
 Items 1.\ and 2.\ shall form the \emph{upper energy surface}.

 \item For $\sigma=-1$ and $m_2 v_2^2 < s <  m_1 v^2_1$, for a domain of parameters $p_n^T$, Eq.\ (\ref{eq:SparamE}) gives minima (\ref{eq:SEclEM}). One boundary of this domain is at $s=m_1 v_1^2$, and the other one is determined by the radius $s_c(u)$, where the second derivative matrix (\ref{eq:SEclEDD}) ceases to be positive definite (its determinant crosses zero). This surface is shown in Fig.\ \ref{fig:qcl}.
 
 \item For $\sigma=-1$ and $s=m_1 v_1^2$, the \emph{Weyl disc} is obtained. Here $p_1^T=0$, and
 \begin{equation}
  \label{eq:SeclWDE}
  E_{{\rm cl},-} = -\frac{m_1 v_1^2}{2} -  \sum_{n\ne 1}\frac{(p_n^T)^2/(2m_n^*m_1^*)}{(v_1/v_n)^2/m_n^* - 1/m_1^*}\,.
 \end{equation}

 \item For $\sigma=-1$ and $s>m_1 v_1^2$, Eq.\ (\ref{eq:SparamE}) yields the minima (\ref{eq:SEclEM}).
 
 Items 4.\ and 5.\ form the \emph{intermediate} and the \emph{lower energy surfaces}, respectively. The solid body on Fig.\ \ref{fig:qcl} shows the parameter values for which both solution exist.
\end{enumerate}

\subsection{Perturbation theory}\label{sec:Spert}
We obtain here formulas for the energy splitting between the degenerate levels in the Weyl disc with the help of perturbation theory. Let us write the Hamiltonian (\ref{eq:Hexc}) in the form
\begin{equation}
 \label{eq:SH}
 \hat{H} = \frac{(\hat{{\bf p}}-{\bf p}^T)^2}{2 m^*} + \hat{H}_{\rm WP} + V({\bf r})\,,
\end{equation}
where we have assumed that the effective masses are isotropic, and $\hat{H}_{\rm WP} = \sum_n v_n \hat{p}_n \hat{\sigma}_n$. Let us note first, that the nanostructure Hamiltonian (\ref{eq:H}) is also of this form, with the replacements $m_n \to (2eP/\hbar)^2 L_n$ ($P$ is an arbitrary constant of momentum dimension), $v_n \to (\hbar/2eP)I_n$, $\hat{p}_n \to P\hat{\phi}_n$, $p^T_n\to P\phi_n^r$ and $x_n \to (\hbar/P)\hat{N}_n$. The potential for the exciton is then $V({\bf r})=e_*^2/(r\pi\epsilon_0 r) + \Delta_{\rm ex}$, and for the nanostructure $V({\bf r}) = \sum_n k_n x_n^2/2 = \sum_n (2e N_n)^2/2C_n$.

We split the Hamiltonian (\ref{eq:SH}) into unperturbed part and perturbation as $H=H_0+H_1$, where $H_1=\sum_{n\ne 1} v_n \hat{p}_n \hat{\sigma}_n$. We start with the solutions of the problem
\begin{equation}
 \label{eq:SH00}
 \left[\frac{{\hat{{\bf p}}}^2}{2m^*} + V({\bf r})\right] \psi_0 = \tilde{E} \psi_0({\bf r})\,.
\end{equation}
In the case of the superconducting nanostructure, these are
\begin{equation}
 \label{eq:Sharm}
 \psi_0(\phi) = \prod_k \phi^H_{nk}(\phi_k)\,,\quad \tilde{E}=\sum_k \hbar \omega_k\left(n_k+\frac{1}{2}\right)\,,
\end{equation}
where $\phi^H_{nk}$ are harmonic oscillator eigenstates with $\omega_k = 1/\sqrt{L_k C_k}$. For the exciton, the eigenfunctions are hydrogen eigenfunctions, and the energy levels are
\begin{equation}
 \label{eq:Shyd}
 \tilde{E}=-\left(\frac{e^2_*}{4\pi\epsilon_0}\right)^2 \frac{m^*}{\hbar^2}\frac{1}{2n^2}\,,
\end{equation}
where $n$ is the principal quantum number, and there is a degeneracy $\ell=0,\dots,n-1$ and $m=-\ell,\dots,\ell$. In two dimensions, $1/2n^2$ shall be replaced by $1/2(n-1/2)^2$, and the degeneracy is due to $m=-n+1,\dots,n-1$. In both cases, $n=1,2,\dots$.

If $\psi_0$ solves Eq.\ (\ref{eq:SH00}), then so does
\begin{equation}\label{eq:Spsi}
 \psi({\bf r}) = \exp\left( \frac{\imagi}{\hbar}\left[(p_1^T-\sigma m_1^* v_1)x_1 + p_2^Tx_2+p_3^Tx_3\right]\right) \psi_0({\bf r})|\sigma \rangle
\end{equation}
solve $H_0 \psi=E_0 \psi$ where $\hat{\sigma}_1 |\sigma\rangle = \sigma |\sigma\rangle$, and $E_0=\tilde{E}-m_1^* v_1^2/2 + \sigma v_1 p_1^T$, or, for the nanostructure $E_0 = \tilde{E} - L_1 I_1^2/2+ \sigma (\hbar/2e)I_1 \phi_1^r$.
Eq.\ (\ref{eq:Spsi}) describes wave functions localized at the two energy minima (i.e., shifted in momentum space).

We obtain the energy splitting in the disc as $\Delta E=2|t|$, where
\begin{equation}
 \label{eq:St}\begin{aligned}
 t&=\langle \sigma | \hat{H}_1 | -\sigma \rangle = \langle \sigma | \sum_k v_k p_k^T \hat{\sigma}_k | -\sigma \rangle \\
  &= \langle \sigma | \sum_k v_k p_k^T \hat{\sigma}_k | -\sigma \rangle_{\rm spin}\int \d^d x \e^{\frac{2\imagi}{\hbar}m_1^*v_1 x_1}|\psi_0({\bf r})|^2
 \end{aligned}
\end{equation}
is the tunneling matrix element between states of the same quantum number localized about the two energy minima.
Evaluating Eq.\ (\ref{eq:St}) for the nanostructure yields
\begin{equation}
 \label{eq:StJ}\begin{aligned}
 t&=    \langle \sigma | \sum_k \hbar I_k \phi_k^r/2e  \hat{\sigma}_k | -\sigma \rangle_{\rm spin}\e^{-2\mathcal{Q}}L_{n_1}(4\mathcal{Q})\,,\\
  &\sim \langle \sigma | \sum_k \hbar I_k \phi_k^r/2e \hat{\sigma}_k | -\sigma \rangle_{\rm spin}\e^{-2\mathcal{Q}} \frac{(4\mathcal{Q})^{n_1}}{n_1!}\,,
\end{aligned}\end{equation}
where $n_1$ is the excitation of the easy direction oscillator. The asymptotic formula holds in the quasi-classical limit, $\mathcal{Q}\to\infty$.

For the exciton, the same kind of calculation for the ground state $\psi_0 = \e^{-\tilde{r}}/\sqrt{\pi}$, where $\tilde{r}=m/\hbar^2 (e^2/4\pi\epsilon_0)r$, yields
\begin{equation}\label{eq:StE}
 t = \frac{\langle \sigma | \sum_k v_k p_k^T \hat{\sigma}_k | -\sigma \rangle_{\rm spin}}{2(1+\mathcal{Q}^2)^2}\sim \frac{\langle \sigma | \sum_k v_k p_k^T \hat{\sigma}_k | -\sigma \rangle_{\rm spin}}{2\mathcal{Q}^4}
\end{equation}
in 3d.
The suppression in Eq.\ (\ref{eq:StE}) is of the power-law type, in contrast to the exponential suppression in Eq.\ (\ref{eq:StJ}), as a result of the non-smooth behavior of the hydrogen wave functions at the origin, owing to the singularity of the Coulomb potential. In 2d, $t=\langle \sigma | p_2^T v_2 \hat{\sigma}_2 | -\sigma\rangle /(1+\mathcal{Q}^2)^{3/2}\sim \langle \sigma | p_2^T v_2 \hat{\sigma}_2 | -\sigma\rangle/\mathcal{Q}^3$.

\subsection{Numerical methods}\label{sec:Snum}
In the case of the superconducting nanostructure, we use the fact, that the Hamiltonian (\ref{eq:H}) contains 3 harmonic oscillators to introduce CAPs as $\hat{\phi}_n-\phi_n^r = (\hat{a}_n{}^\dagger + \hat{a}_n)/\alpha_n$, yielding
\begin{equation}
 \label{eq:SHCAP}
 \hat{H} = \sum_n \hbar \omega_k \left(\hat{a}_n{}^\dagger\hat{a}_n+\frac{1}{2}\right) + \sum_n\frac{\hbar I_n}{2e}\hat{\phi}_n\hat{\sigma}_n\,,
\end{equation}
where $\omega_n = 1/\sqrt{L_n C_n}$, $\alpha_n = e\sqrt{2/\hbar}(L_n/C_n)^{1/4}$, and $[\hat{a}_n,\hat{a}_m{}^\dagger] = \delta_{nm}$.

The quasi-classical analysis presented in the main text yields that the ground state is centered at $\phi_1=\pm \phi_0$, $\phi_{2,3} =0$, similarly to a coherent state with parameter $\pm \sqrt{Q}$, therefore the number of states necessary to expand it is $N\sim \mathcal{Q}$. In the orthogonal directions, we use $N/2$ states, and 2 for spin. We find he low-lying eigenvalues of the resulting sparse $N^3/2\times N^3/2$ matrix with Julia library routines. For Fig.\  \ref{fig:junction}, we have used $N=50$.

In the case of the exciton, we use a standard finite difference approach, with a 5 point stencil for both second derivatives, and Richardson extrapolation from two different grid spacings to enhance the accuracy. The grid used is equidistant in a logarithmic variable to enhance resolution in the vicinity of the origin, where the wave function is non-smooth. We use a couple of hundred points in each direction. The accuracy is verified with computing the standard Coulomb eigenvalues on the same grid, and comparing them to the exact results. For reproducibility we have published the code on Zenodo repository \cite{TheCode}.


\end{document}